\documentclass[12pt,preprint]{aastex}

\shorttitle{X-ray eclipse of the AGN in NGC 1365}
\shortauthors{G. Risaliti et al.}

\begin{document}

\title{Occultation measurement of the size of the X-ray
emitting region in the Active Galactic Nucleus of NGC
1365
}

\author{
G. Risaliti\altaffilmark{1,2}, M. Elvis\altaffilmark{1}, 
G. Fabbiano\altaffilmark{1}, A. Baldi\altaffilmark{1},
A. Zezas\altaffilmark{1},
M.~Salvati\altaffilmark{2}
} 
\email{grisaliti@cfa.harvard.edu}

\altaffiltext{1}{Harvard-Smithsonian Center for Astrophysics, 60 Garden St. 
Cambridge, MA 02138 USA}
\altaffiltext{2}{INAF - Osservatorio di Arcetri, L.go E. Fermi 5,
Firenze, Italy}
\begin{abstract}
We present an occultation of the central X-ray emitting region in the
Seyfert Galaxy NGC 1365. This extreme spectral variation (from Compton-thin
to reflection-dominated and back to Compton-thin in four days) has been
caught in a ten days {\em Chandra} monitoring campaign consisting of six short (15~ks)
observations performed every two days. We discuss the implications of
this occultation within the scenario of a Compton-thick cloud 
crossing the line of sight of the X-ray source. We estimate a source
size $R\leq10^{14}$~cm and a distance of the cloud from the source
$D\leq10^{16}$~cm. This direct measurement confirms the theoretical expectations
of an extremely compact X-ray source, and shows that the Compton-thick circumnuclear
gas is located at a distance from the center on the scale of the Broad Line Region. 
\end{abstract}

\keywords{ Galaxies: AGN --- Galaxies: individual (NGC 1365)}

\section{Introduction}

Variability of the X-ray absorbing gas is common in Active Galactic Nuclei
on time scales from months to years 
(Risaliti, Elvis \& Nicastro~2002). If an observed variation
of absorbing column density $N_H$ is due to clouds in virialized motion
crossing the line of sight,
the amount and the duration of the $N_H$ variation 
put constraints on the distance of the obscuring cloud from the center,
and on the density of the cloud.

In a few cases an $N_H$ variation has been detected within a single observation
(NGC~4388 Elvis et al.~2004, NGC~4151, Puccetti et al.~2007),
indicating that the absorber must be extremely compact, i.e. on the scale of
or slightly larger than the Broad Line Region ($\sim10^{16}$~cm for a $10^8~M_\odot$
black hole).  

In this context, the Seyfert Galaxy NGC~1365  has shown extreme variability
in the past $\sim12$ years: it was observed in a reflection-dominated
state by ASCA in 1995 (Iyomoto et al.~1997), then in a Compton-thin state by
BeppoSAX in 1997, with absorbing column density $N_H\sim4\times10^{23}$~cm$^{-2}$
(Risaliti et al.~2000). Such a long time interval between the observations leaves
two possible scenarios open: extreme absorption variability ($\Delta(N_H)>10^{24}$~cm$^{-2}$)
or a switch off and on of the X-ray source.
This ambiguity has been solved with a more recent set of short observations, performed
by {\em Chandra} in December 2002, and by {\em XMM-Newton} three and six weeks later.
The source was caught in a reflection-dominated state in the first and third observation,
while it was in a Compton-thin state with $N_H\sim4\times10^{23}$~cm$^{-2}$
 in the second observation (Risaliti et al.~2005A, hereafter R05). Such fast variations 
are hard to explain within the intrinsic variation scenario strongly suggesting that 
the observed reflection-dominated states are due to  Compton-thick clouds crossing
the line of sight (see R05 for a full discussion). 
Three additional {\em XMM-Newton} observations performed in
2003 and 2004 caught the source in 
a Compton-thin state, with column densities between 1.5 and 5$\times10^{23}~$cm$^{-2}$.

The latest two observations are relatively long (60~ks) and allowed a detailed timing
and spectral analysis. The spectral analysis revealed the presence of a highly ionized,
compact absorber (Risaliti et al.~2005B), while the timing analysis revealed
column density variability of $\Delta(N_H)\sim10^{23}$~cm$^{-2}$ on time scales of $\sim50$~ks
(Risaliti~2006, Risaliti et al.~2007, in prep.).

In order to explore the variability time scales between the longest single observations 
($\sim$a day) and the shortest observed Compton-thick/Compton-thin change (three weeks)
we conducted a {\em Chandra} campaign consisting of six 15~ks observations
performed once every $\sim2$~days for ten days in April~2006.
Here we report the results of these observations, with emphasis on an occultation
event which occurred during the first four days of monitoring, and we discuss the physical
implications of these results. 
\section{Reduction and Data Analysis}

The observation log is shown in Tab.~1. All the observations were performed with the
ACIS-S instrument (Weisskopf et al.~2002) in '1/4 window' mode in order to avoid possible pile-up. 
A check of the reduced spectra confirmed that in all cases the pile-up is lower than 1\%.

The data were reduced using the CIAO~3.3\footnote{http://asc.harvard.edu/ciao/}
package and using a standard
procedure, as described in the CIAO Threads$^1$. 
We extracted the spectrum from a circular region with a 2~arcsec radius.
This removes most of the soft, diffuse emission from the spectra (see Fig~1 in R05). 
A complete analysis of the spectral and spatial properties of the diffuse component will be
presented elsewhere (Bianchi et al.~2007, in prep.). 
The background was selected from a region in the field free from contaminating
sources. The spectral analysis was performed using the SHERPA package inside CIAO.
\begin{table}
\caption{NGC 1365 - New Chandra Observations}
\centerline{\begin{tabular}{lcccc}
\hline
OBS & Delay$^a$ & Duration & Cts/s  & Cts/s \\
    &  (ksec)      & (ksec)          &  (0.5-3 keV)  & (3-10 keV) \\  
\hline
OBS 1 & 0   & 13.86 &3.6$\time10^{-2}$  & 6.1$\times10^{-2}$\\
OBS 2 & 178 & 15.16 &3.3$\time10^{-2}$  & 2.2$\times10^{-2}$\\
OBS 3 & 197 & 15.18 &3.5$\time10^{-2}$  & 5.6$\times10^{-2}$\\
OBS 4 & 228 & 15.15 &3.2$\time10^{-2}$ & 10.2$\times10^{-2}$\\
OBS 5 & 212 & 16.12 &3.4$\time10^{-2}$ & 10.5$\times10^{-2}$\\
OBS 6 & 229 & 15.09 &3.3$\time10^{-2}$ & 15.9$\times10^{-2}$\\
\hline
\end{tabular}}
Observation log of the six {\em Chandra} ACIS-S observations,
started on April~14, 2006. $^a$: Time elapsed from the end of the previous observation.
\end{table}

A simple visual inspection of the spectra (Fig.~1) is enough to notice the main result
of this work: the second spectrum is completely different from the others
both in flux (it is much fainter) and in shape (it is flatter, with a prominent
emission feature at $\sim6.5$~keV), indicating that the source switched to a reflection-dominated
state between the first and second observation, and then switched back to
a transmission-dominated state between the second and the third observation.

We first performed a simple spectral analysis, 
with a model consisting of a thermal
emission at low energies, an absorbed power law and an iron emission line 
with E=6.4keV. In all the fits discussed below, the soft component is 
fitted with a thermal component (Raymond \& Smith~1977) 
with temperature kT=0.8~keV and constant within 2\% in
all cases. 

All the high energy (E=2-10~keV) spectra except 
for the second one are well fitted (reduced $\chi^2$
between 1.1 and 1.5) with a
power law with photon index $\Gamma=1.8-2$,
a column density $N_H=2-4\times10^{23}$~cm$^{-2}$ and an iron line with equivalent
width EW=200-300~eV.
The second spectrum is fitted by a flat power law with $\Gamma\sim0.5$ and
$N_H<10^{21}$~cm$^{-2}$. The emission feature has an 
equivalent width EW$\sim1.5$~keV.

These results confirmed the visual inspection, and prompted a more detailed analysis, which 
has been performed in two main steps:\\
1) We fitted the second spectrum with a cold reflection continuum 
(PEXRAV model, Magzdiarz \& Zdziarski~1995) and an emission
line at 6.4~keV, corresponding to neutral iron K$\alpha$ emission. 
A second line at energy E=6.9~keV, corresponding to hydrogen-like iron, is also
required by the fit. The photon index of the intrinsic component is not
well constrained, so we fixed it to the average of the values obtained fitting
the transmission-dominated spectra. The best fit equivalent widths of the two lines are
EW$_{6.4}=1.2\pm0.2$~keV and EW$_{6.9}=0.9\pm0.4$~keV. \\
2) The other five spectra were fitted with the same reflection continuum (with all
the parameters frozen to the best fit values obtained in the analysis of the second
spectrum) plus a continuum component representing the direct emission of the X-ray source.
Since the spectra obtained in the past
{\em XMM-Newton} observations have a much higher signal-to-noise, we 
used the best fit models obtained from those observations as a baseline for our spectral
fitting. 
The best fit model consists of an absorbed power law, plus 
four narrow absorption lines between 6.7 and 8.3 keV (representing He-like and
H-like iron Fe K$\alpha$ and K$\beta$ transitions), and a broad emission
line.
We refer to Risaliti et al.~2005B for a detailed description of these models.

The results of the fits are in all cases satisfactory from a statistical point of view
(reduced $\chi^2$ between 1 and 1.2) and provide best-fit values for the continuum photon
index constant within the errors, while the values of the column density show
significant variations. We then repeated the whole analysis fitting all the spectra
simultaneously, requesting a constant value for the photon index and leaving the other
parameters free to vary. We obtained as good a fit as in the previous case from a
statistical point of view (overall reduced $\chi^2$=1.01) and slightly smaller error intervals for
the column density estimates. As a side product of our analysis, we mention that the two
strongest iron absorption lines ($K_\alpha$ lines from He-like and H-like iron) are significantly
detected. The inclusion of these features in the fit is relevant in our context 
only because
neglecting them would affect the estimate of the continuum parameters. However, their
detection is an interesting independent confirmation of the highly ionized absorber
discovered in the XMM-Newton observations (Risaliti et al.~2005B). 
We will discuss this issue elsewhere
(Risaliti et al. 2007, in prep.).

The main best fit parameters are shown in Table~2.
\begin{figure}
\epsscale{0.9}
\plotone{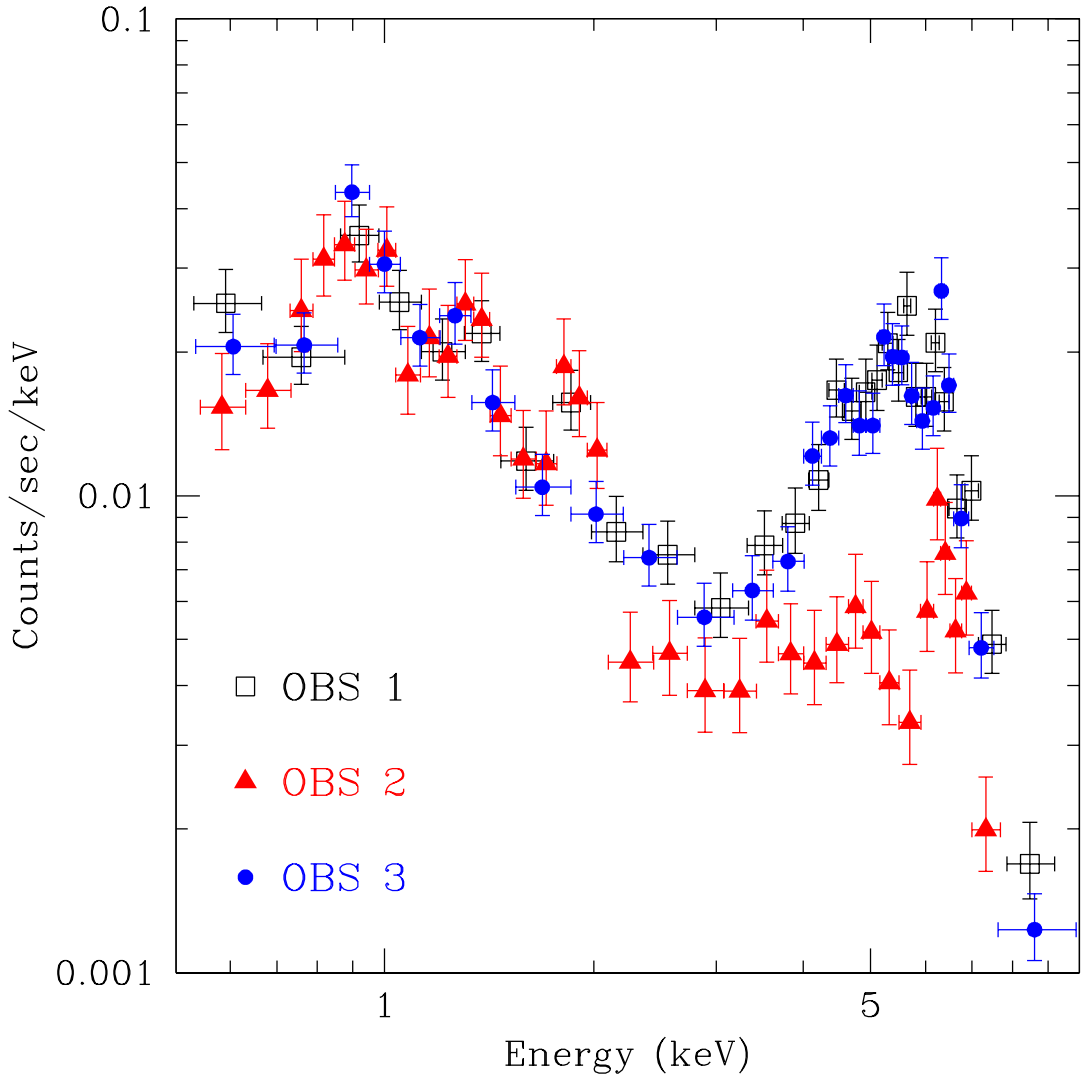}
\caption{
Spectra obtained from the first three {\em Chandra} observations of NGC~1365. 
the time
interval between the single observations is two days.
The first and third spectrum are typical of a transmission-dominated source,
with a steep continuum and a photoectric cut-off at $\sim4$~keV.    
The second spectrum is fainter by a factor $\sim10$, and is characterized 
by a flat continuum, and a prominent iron
emission line, typical of reflection-dominated sources. The last three 
{\em Chandra} observations are similar to the first and third ones, and are
not shown for clarity.
}
\end{figure}

\begin{figure}
\epsscale{0.8}
\includegraphics[width=12.5cm,angle=-90]{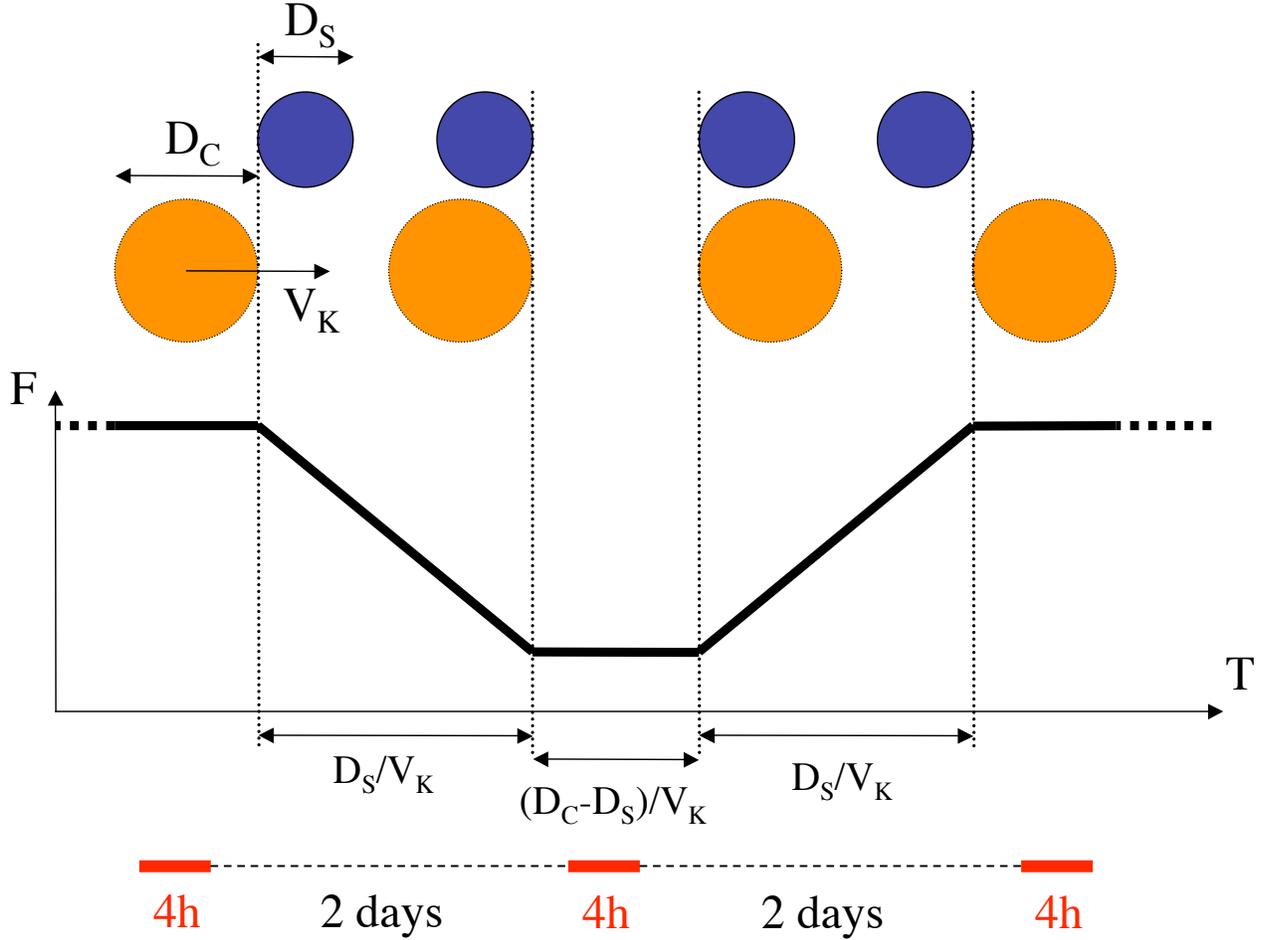}
\caption{Schematic representation of the observed eclipse. The intervening thick 
cloud (orange circle, diamater $D_C$) vith Keplerian velocity $V_K$ starts covering the 
X-ray source (blue circle, diameter $D_S$) at some time between the first and 
second Chandra observations. After a time  $T_1$= $D_S/V_K$ the source is completely 
covered, and remains obscured for a time $T_2=(D_C-D_S)/V_K$. In this state it is
observed for the second time by Chandra. Then, it gradually uncovers, until it is
back in the initial state. After some more time it is observed again by Chandra.
From the times between the observations we infer $T_2+2T_1 < 4$~days; $T_2 > 4$~hours.
The lower part of the figure shows the time evolution of the observed
flux.
}
\end{figure}
\begin{table}
\caption{NGC 1365 - Spectral fits}
\centerline{\begin{tabular}{lccccc}
\hline
OBS & $\Gamma$ & $N_H^a$ & $N_{H,2}^b$ & $A^c$ &R$^d$\\
\hline
OBS 1 & 1.4$^{+1.1}_{-0.4}$ & 46$^{+11}_{-12}$ & 40$^{+6}_{-3}$ & 3.7$^{+0.8}_{-0.7}$ & 0.7$\pm0.2$\\
OBS 2 & 1.7$^d$                 & $>100$           & $>100$     & 2.6$^{+0.8}_{-0.9}$ & --\\
OBS 3 & 2.7$^{+1.5}_{-1.0}$ & 60$^{+19}_{-13}$ & 49$^{+6}_{-6}$ & 3.1$^{+0.7}_{-0.6}$ &0.8$\pm0.2$\\
OBS 4 & 1.8$^{+0.7}_{-0.6}$ & 36$^{+7}_{-7}$   & 34$^{+3}_{-3}$ & 4.3$^{+0.5}_{-0.5}$ &0.6$\pm0.1$\\
OBS 5 & 2.0$^{+0.5}_{-0.6}$ & 44$^{+7}_{-7}$   & 41$^{+3}_{-3}$ & 5.6$^{+0.7}_{-0.6}$ &0.5$\pm0.2$\\
OBS 6 & 1.4$^{+0.4}_{-0.2}$ & 22$^{+3}_{-2}$   & 23$^{+1}_{-1}$ & 4.6$^{+0.2}_{-0.3}$ &0.6$\pm0.1$\\
\hline
\end{tabular}}
$^a$: Column density in units of 10$^{22}$~cm$^{-2}$, obtained fitting a model with a
free photon index power law.
$^b$: Same, with the photon index frozen to the average value, $<\Gamma>=1.68$.
$^c$: Normalization of the power law, in units of 10$^{-3}$ Photons~s$^{-1}$~cm$^{-2}$~keV$^{-1}$.
In OBS~2 this refers to the intrinsic emission producing the observed reflected spectrum.
$^d$: Fixed value. $^d$: ratio between the normalizations of the reflection and transmission
components.
\end{table}

\section{Discussion}
The analysis described in the previous Section shows that a change from
Compton-thin to Compton-thick states, and then back to Compton-thin,
occurred in the first four days of our {\em Chandra} monitoring of NGC~1365.

As discussed in R05, such rapid changes are hard to explain with intrinsic
spectral variations. In order to further check this possibility, we estimated
the upper limit of a possible direct component in the reflection-dominated spectrum.
Assuming a spectral shape analogous to that obtained for the first
and third observations (which are quite similar both in flux and spectral parameters, Table~2)
the 90\% upper limit to the direct flux is only 5\% of that of observations one and three. 
A decrease of the intrinsic flux by a factor of 20, and then an increase back to the initial
flux, besides implying an unlikely "fine tuning" in order to reproduce the observed symmetry between
the fading and recovery phase, would require a cooling time of at least 3-4~weeks in the
framework of a Shakura-Sunyaev (1973) disk (see R05 for details), completely incompatible with our
observed varation times. For this reason, in the following we will only discuss the occultation
scenario. 

In the simplest scheme, shown in Fig.~2, the size of the X-ray emitting source $D_S$ is given by the
obscuring cloud velocity, $V_K$, times the ingress/egress time $T_1$. The linear size of the
obscuring cloud is $D_C>D_S$ and the distance between the cloud and the source is R. Since
our goal is to put an upper limit on the source size, it is particularly important to discuss
the upper limits on the estimates of these two parameters. 

Here we discuss several constraints on $V_K$, $T_1$ and $D_S$.\\
{\bf A. Statistical limits on $T_1$}. 
The occultation event
is characterized by two times (Fig.~2): the ingress/egress time $T_1=D_S/V_K$ and the time during
which the source remains completely obscured, $T_2=(D_C-D_S)/V_K$. Considering the X-ray
observational history of NGC 1365, we note that the source has been observed four
times in a completely reflection-dominated state, while it has never been observed 
during the ingress/egress phase, in any published observation. In order to check this, 
we re-analyzed all the past observations looking for (a) fast drops in the light curves of
the transmission-dominated spectra, indicative of a possible occultation event during the
observation, and (b) possible direct continuum components in the reflection-dominated 
spectra. The first check easily ruled out the possibility that such occultations happened
during any observation (this would imply a drop by a factor of at least 10 in the hard X-
ray (E$>$4 keV) light curve, which would be easily detected). In the second check we 
added a direct continuum component to the model of the reflection-dominated spectra,
consisting of an absorbed power law. We required that the photon index and the
absorbing $N_H$ varied within the lowest and highest values measured in all the 
transmission-dominated spectra. In the three most recent reflection-dominated
spectra this extra component is not required, and the upper limit to the flux of this 
component is as low as 10\% of the faintest transmission-dominated spectrum. The result 
for the ASCA observation is inconclusive, because of the lower signal-to-noise. Also in 
this case the extra component is not statistically required, but a significant contribution
cannot be ruled out. Therefore, we do not include this observation in our analysis.
Summarizing, in three cases the source was found in a completely reflection dominated
state, and in no cases it was caught during an ingress/egress. During an eclipse the 
source is in the partial occultation phase (state 1) for a total time $2T_1$, and is totally
covered (state 2) for a time $T_2$. The fraction of time in the completely obscured state is
$f=T_2/T_{TOT}$ where $T_{TOT}=(2T_1+T_2)<4.2$~days based on our observations. We require that 
the probability of finding the source three times in state~2 but never 
in state~1 is P=$f^3>$10\%. This implies
$T_2>0.46 T_{TOT}$ and $T_1<0.27\times T_{TOT} < 1$~day. Furthermore, we get 
$T_2>1.7 T_1$, and $D_C>2.7 D_S$.\\ 
{\bf B. Physical limits on $V_K$}. A first limit on $V_K$ comes from 
the measured width of the iron emission line in the reflection-dominated spectra. Here we assume
that the reflecting Compton-thick gas is the same one which is 
responsible for the occultations. This is in agreement with the available statistics:
since during the whole observational history of the source the occurrence of Compton-thick
states is 4/14, it is expected that the obscuring clouds cover a significant fraction,
$\sim$1/3, of the solid angle as seen from the source, and therefore they are also expected to
contribute significantly to the observed reflection (a typical fitting value of the
PEXRAV parameter R, normalized to 1 for a half-solid angle coverage of the reflector,
is indeed $\sim0.6-0.8$, as shown in Tab.~2). 
The best available estimate of the iron line width, $W_{Fe}$, comes from the XMM-Newton
observation of the reflection-dominated state of NGC 1365. We obtain $W_{Fe}<150$~eV, 
corresponding to a velocity $V<$7,000~km/s. The line width measures only the average  
line-of-sight velocity, while during the occultation the cloud velocity vector lies in the
plane of the sky. Assuming circular orbits, this implies that the actual transverse
velocity during occultation can be as large as $(\pi/2)\times V=$12,000~km/s. Finally,
using $T_1<1$~day we obtain $D_S < 10^{14}$~cm.\\
{\bf C. Geometrical limits on $D_S$}. A geometrical limit on $D_S$ can be obtained assuming the
minimum possible distance R between the source and the cloud. This is given by  
$R=(D_S+D_C)/2$. Since the cloud size $D_C$ must be equal to or larger than the source size
$D_S$, the minimum distance for a given $D_S$ is $R_{min}=D_S$. If the cloud is moving with
Keplerian velocity $V_K=(GM_{BH}/D_S)^{1/2}$ the condition $V_K\times T_1=D_S$ implies 
$D_S=(GM_{BH})^{1/3}T_1^{2/3}$. 

Two independent estimates of the black hole mass in NGC
1365 are available: log$(M_{BH}/M_\odot)=7.3\pm0.4(0.3)$ from the $M_{BH}$-bulge velocity dispersion
correlation (Oliva et al.~1995, Ferrarese et al.~2006) and log$(M_{BH}/M_\odot)=7.8\pm0.4(0.3)$ 
from the relation between $M_{BH}$ and the
K magnitude of the host bulge (Dong \& De~Robertis~2006, Marconi \& Hunt~2003)
 where the errors include statistical and systematic
effects, and the number in brackets refer to the statistical dispersion of the correlation.

Using $T_1<1$~day we obtain $D_S<3\times10^{14} (M_{BH}/M_{BEST})^{1/3}$~cm,
where
$M_{BEST}$ is the average estimate of the black hole mass according to the correlations cited
above. 
Expressing the source size in units of gravitational radii we 
find $D_S<33(M_{BH}/M_{BEST})^{-2/3} R_G$.
We note that the higher the black hole mass, the larger the physical size of the source,
and the smaller the source size in units of gravitational radii. Considering the relatively
high uncertainty in the black hole mass determination, we can repeat the above
calculations using the minimum and maximum values compatible with the two
correlations, log$M_{BH(min)}=7.1$ and log$M_{BH(max)}=8.0$. We obtain
$D_S(min)<2\times10^{14}$~cm, corresponding to 56~$R_G$, and $D_S(max)<4.6\times10^{14}$~cm, corresponding to
15~$R_G$.\\
{\bf D. Physical limits on R}. A conceptually different limit on the distance R between the
source and the obscuring cloud can be obtained by considering the ionization state of the
latter. This is defined by the ionization parameter, $U=L_X/(nR^2)$, where $n=N_H/D_C$, is the
density of the obscuring cloud, located at a distance R from the central source,
whose X-ray total luminosity is $L_X$. A limit on the ionization parameter can be
obtained from the analysis of the reflection-dominated spectra observed with XMM-
Newton and Chandra: we added to the best fit model an extra component consisting of
the continuum observed in transmission-dominated states, absorbed by gas with variable
column density and ionization parameter (Done et al.~1992). 
Since the upper limit on the flux of the direct
emission is only a few percent of that observed in transmission-dominated states, the
absorber must be effective enough to remove it almost completely. This implies a lower
limit on the absorbing column density $N_H>10^{24}$~cm$^{-2}$ and an upper limit on the
ionization parameter $U<U_{max}=100$. From the latter limit, assuming that the cloud is moving
with Keplerian velocity $V_K$ and requiring that the cloud dimension $D_C$ is larger than the
source dimension $D_S$ we obtain, after a little algebra: 
$R>(GM_{BH})^{1/5}[(T_1+T_2)L_X/(U_{max}N_H)]^{2/5}$. Assuming fiducial values for the 
occultation times and the cloud column density ($T_1+T_2\sim2$~days and $N_H=10^{24}$~cm$^{-2}$)
we obtain $R>3\times10^{15} (M_{BH}/M_{BEST})^{2/5}$~cm. 
We note that the dependence on the
black hole mass is weak, so even adopting the extreme values allowed for $M_{BH}$
would change the result by a factor of less than 30\%. From this estimate
we can easily obtain a new limit on the Keplerian velocity $V_K=(GM_{BH}/R)^{1/2} < 12,000
(M_{BH}/M_{BEST})^{2/5}$~km/s, i.e. the same upper limit as using the previous argument on the line
width. Analogously, the limit on the source size is $D_S=V_KT_1<10^{14}(M_{BH}/_{MBEST})^{2/5}$~cm.

Summarizing the constraints on the source size, we obtained:\\
1)      $D_S<3\times10^{14} (M_{BH}/M_{BEST})^{1/3}$ cm (from geometrical considerations)\\
2)      $D_S<10^{14}$ cm (from the observed limits on $V_K$)\\
3)      $D_S<10^{14} (M_{BH}/M_{BEST})^{2/5}$ cm (from limits on the ionization state of the obscuring
cloud)\\
All these independent methods strongly suggest that the X-ray source has a size of the
order of or smaller than $10^{14}$~cm, corresponding to 10~$R_G$ for a black hole mass
$M_{BH}=M_{BEST}=3\times10^7 M_\odot$.

We note that our estimates can be slighly altered if the case of a partial covering of the thick cloud
in the first and third observations is considered. Specifically, in order to reproduce the observed symmetry
in the first three observations, the same fraction F of the source should be covered in observations one
and three. However, if the fraction F is small, the effect is negligible, while if F is big,
the cloud would uncover the source in the subsequent observations, implying 
a large increase of the flux from the third to the fourth observations,
which is not observed (Table~2). We conclude that the possible effect of partial covering of
the transmission-dominated spectra would not significantly affect our estimates of the 
source and cloud sizes.

Using the calculations discussed above we can easily estimate the distance of the obscuring
cloud from the center.
Assuming $V_K=$12,000~km/s and $M_{BEST}=3\times10^7 M_\odot$ we obtain $R=3\times10^{15}$~cm,
compatible with the lower limit deduced from the ionization condition.
Such a distance corresponds to ~300~$R_G$, and is of the same order of the distance 
from the center of the innermost broad line clouds, and therefore much more compact than 
the circumnuclear medium assumed in the standard Unified Models of AGNs (e.g. Krolik \& Begelman~1988).

\acknowledgements
This work has been partially funded by NASA Grant G06-7102X. We also acknowledge financial
contribution from contract ASI-INAF I/023/05/0. 


\end{document}